\documentclass[12pt]{iopart}
\usepackage{dsfont}
\def\one{\mathds{1}}
\begin{document}

\bibliographystyle{unsrt}

\title{Simple derivation of the special theory of relativity
without the speed of light axiom}

\author{O Certik$^{1,2}$}

\address{$^1$ Faculty of Mathematics and Physics, Charles University, Ke
Karlovu 3, 121 16 Praha, Czech Republic}
\address{$^2$ Institute of Physics, Academy of Sciences of the Czech
Republic, Cukrovarnick\'a 10, 162 53 Praha 6, Czech Republic}
\ead{certik@fzu.cz}
\begin{abstract}
We show a very simple yet rigorous derivation of the invariance of the
space-time interval (and hence the whole special relativity)
just from the isotropy, homogeneity and a principle of relativity, without the need
of the speed of light axiom. This article is intended as a textbook explanation
of the special relativity.

\submitto{\EJP}
\end{abstract}

\maketitle

\section{Introduction}

It is well-known, that the special theory of relativity can be derived
without the speed of light axiom, see for example \cite{Pal},
\cite{sonego-pin}, \cite{torretti}, \cite{coleman} and follow the citations in
these papers, some of them dating back to Kaluza 1924 and Ignatowski 1910.
However, the standard texts on the special and general relativity like \cite{schutz},
\cite{goldstein}, \cite{feynman}, \cite{misner}, \cite{weinberg} don't mention
this at all. Only \cite{dixon}, \cite{torretti} and \cite{rindler}
discuss this issue, but except for the last one, the treatment is still too much complex.

It was reported that the possibility of deriving the special relativity
without the speed of light axiom was discovered many times in the past
\cite{sonego-pin},
without realizing that this was already clear to Albert Einstein at the
time of the writing \cite{einstein}.

The best argument on this issue is given in \cite{rindler}: either particles
can be accelerated to arbitrary speeds, or they cannot. If they can, we get the
Galileo transformation, if they cannot, then there must exist, mathematically
speaking, a least upper bound $c$ to particle speeds in any one inertial frame.
By the relativity principle, this bound must be the same in all inertial
frames, moreover, the speed $c$ -- whether attained or not by any physical
effect -- must transform into itself (otherwise we could get higher speed than
$c$ of some particle when transformed from $S$ to $S'$). But when $c$
transforms into itself, we are lead uniquely to the Lorentz transformation by
the usual procedure employed in most of the texts. Thus the relativity
principle by itself necessarily implies that all inertial frames are related
either by Galilean transformations, or by Lorentz transformations with some
universal $c$. The only role of the speed of light axiom is the determination
of $c$.

However it is not really intuitive that particles cannot be accelerated to
arbitrary speeds. On the other hand, the fact that we will
allow any possible transformation between S and S' (and derive the only two
allowed possibilities) is much more plausible. It is of course equivalent,
but the latter approach is more explicit.

The basic principles which the Newtonian theory (and also the special theory of
relativity) is built on are homogeneity, isotropy and the principle of
relativity. This allows two and only two possible transformations: Galilean and
Lorentzian.  Experimentally the Galilean is not satisfactory for many reasons
(the apparent speed of light limit and other problems), so we need to take the
Lorentz one.  There is no other option left, unless we want to sacrifice the
principle of relativity or homogeneity or isotropy.

Almost every aspect of this issue can already be found in the literature.
However, what the author couldn't find, is a derivation of the
special relativity in a rigorous, but simple, short and clear way. The amount of
rigor is subjective, also some of the assumptions can be weakened, or made more
precise, but what we want to achieve in this article is to choose some small
amount of assumptions, put them into equations and from that point only work
with the algebra. See the references, for example \cite{field1}, \cite{field2}
for a thorough description of what postulates are necessary and which can be
weakened and also for a review of all the derivations of the Lorentz
transformations known to the author of \cite{field1}, \cite{field2} until 1997
(together with his own new derivation --- but we present a shorter one in this
paper).

Many articles (see the citations in \cite{Pal}) first derive the
velocity addition law and the arguments are quite messy, referring to pictures
many times \cite{coleman}, or nitpicking in unnecessary mathematics
\cite{torretti}, \cite{gorini3}, \cite{gorini4}, \cite{matolcsi} etc. The
articles
\cite{gorini1}, \cite{gorini2} are very good and cover almost everything which
is shown in this article, however they also concentrate on quite unimportant
mathematical details and some of their derivations are unnecessarily
complicated and long. The best approach known to the authors is \cite{Pal}
that derives the Lorentz transformation and the velocity addition law using a
very clear arguments, first writing down algebraic relations that are
equivalent to homogeneity, isotropy and the relativity principle and the rest
is a pure algebra. He works in 2D spacetime though and only derives the Lorentz
boost.

In this paper we try to use the same, nice and simple arguments of \cite{Pal},
but using the results from \cite{gorini1}, \cite{gorini2} and \cite{torretti},
thus deriving everything in 4D spacetime and not only showing how to get the
Lorentz boost, but also that all permissible transformations obey the
orthogonal property, thus proving the invariance of the spacetime interval. And
it is well-known, that the whole special theory of relativity can be derived
from the invariance of the interval.

\section{Derivation of the transformation}

This short section is the main result of the article. The other sections are
just more detailed explanations and discussions.

Let's have two Cartesian systems $S$ and $S'$, where $S'$ is moving with the
velocity $v$ along the $x$-axis and at the time $t=0$, $S=S'$ (in other words
the $y$ and
$z$ axes of both systems are parallel and the $x$ axes are the same, except
that the origins $x=0$ and $x'=0$ are moving with the speed $v$ with respect to
each other: when $x=vt$, then $x'=0$).

We need to assume the homogeneity, isotropy and the principle of relativity.
In sections \ref{homogeneity}, \ref{relativity}, \ref{isotropy} we show in
detail, that these very general and "obvious" assumptions can be written
mathematically using the following equations \eref{ass-1}-\eref{ass-5} (if some
of them look unintuitive or confusing, look into the sections \ref{homogeneity},
\ref{relativity}, \ref{isotropy} for the thorough derivation and explanation):

$$
\left( \begin{array}{c}
t' \\
x' \\
y' \\
z' \\
\end{array} \right)
=
A(v)
\left( \begin{array}{c}
t \\
x \\
y \\
z \\
\end{array} \right)
$$
where A(v) is a matrix
\numparts
\begin{equation}
A(v)=
\left( \begin{array}{cccc}
a_{00} & a_{01} & a_{02} & a_{03} \\
a_{10} & a_{11} & a_{12} & a_{13} \\
a_{20} & a_{21} & a_{22} & a_{23} \\
a_{30} & a_{31} & a_{32} & a_{33} \\
\end{array} \right) \label{ass-1}
\end{equation}
and the coefficients $a_{\mu\nu}$ only depend on $v$ (homogeneity). We require
\begin{equation}
A(0) = \one\label{ass-7}
\end{equation}
and also (relation between origins and parallel axes)
\begin{eqnarray}
x'=0          &\quad\mbox{when $x=vt$, $y=0$, $z=0$}\label{ass-9}\\
x'=0\quad y'=0&\quad\mbox{when $x=0$,  $y=0$, $z$ arbitrary}\label{ass-10}
\end{eqnarray}
For each $v$ (relativity):
\begin{equation}
A(-v)A(v) = \one \label{ass-2}
\end{equation}
For each $u$ and $v$ there exist $w$ such that (relativity):
\begin{equation}
$$A(u)A(v) = A(w)$$ \label{ass-3}
\end{equation}
For each $v$ there exist $\bar v$ such that (isotropy)
\begin{equation}
$$TA(v)T = A(\bar v)$$ \label{ass-4}
\end{equation}
where the matrix $T$ is
$$T=
\left( \begin{array}{cccc}
1 & 0 & 0 & 0 \\
0 & -1 & 0 & 0 \\
0 & 0 & 1 & 0 \\
0 & 0 & 0 & 1 \\
\end{array} \right)
$$
For each $v$ and each $\alpha$ (isotropy):
\begin{equation}
$$R(-\alpha)A(v)R(\alpha) = A(v)$$ \label{ass-5}
\end{equation}
\endnumparts
where the matrix $R(\alpha)$ is:
$$
R(\alpha) =
\left( \begin{array}{cccc}
1 & 0 & 0 & 0 \\
0 & 1 & 0 & 0 \\
0 & 0 & \cos\alpha & \sin\alpha \\
0 & 0 & -\sin\alpha & \cos\alpha \\
\end{array} \right)
$$

In \ref{deriv} it is shown, that by a pure algebraic manipulation, the above
assumptions directly imply that
$$
A(v)=
\left( \begin{array}{cccc}
{1\over\sqrt{1-Kv^2}} & -{Kv\over\sqrt{1-Kv^2}}& 0 & 0 \\
-{v\over\sqrt{1-Kv^2}}& {1\over\sqrt{1-Kv^2}} & 0 & 0 \\
0 & 0 & 1 & 0 \\
0 & 0 & 0 & 1 \\
\end{array} \right)
$$
where $K$ is an arbitrary constant independent on $v$. This is the Lorentz
($K>0$) and Galilean ($K=0$) transformation.

Let us first review the equations above to see that they really are what we
mean by the homogeneity, isotropy and the principle of relativity.  And then
we'll discuss the above result more thoroughly.

\section{Assumptions}

\subsection{Homogeneity\label{homogeneity}}

The most general transformation from $S$ to $S'$ is:
\begin{eqnarray*}
t' &= T(t,x,y,z,v)\\
x' &= X(t,x,y,z,v)\\ 
y' &= Y(t,x,y,z,v)\\ 
z' &= Z(t,x,y,z,v)
\end{eqnarray*}

The length of a rod put on the $x$-axis in the frame $S$ is
$$l = x_2-x_1$$
and in the frame $S'$ the length will generally be different:
$$l' = x_2'-x_1' = X(t, x_2, 0, 0, v) - X(t, x_1, 0, 0, v)$$
Homogeneity means, that if we move the left end of the rod in the frame $S$
from $x_1$ to $x_1+h$, the right end will move to $x_2+h$ giving the same
length $l = (x_2+h) - (x_1+h) = x_2 - x_1$ and that in the frame $S'$ the
new length $l'=X(t, x_2+h, 0, 0, v) - X(t, x_1+h, 0, 0, v)$ will also be the
same
as before: 
$$X(t, x_2, 0, 0, v) - X(t, x_1, 0, 0, v) = X(t, x_2+h, 0, 0, v) - X(t, x_1+h,
0, 0, v)$$
so
$$X(t, x_2+h, 0, 0, v) - X(t, x_2, 0, 0, v) = X(t, x_1+h, 0, 0, v) - X(t, x_1,
0, 0, v)$$
and dividing by $h$ and taking a limit $h\to0$:
$$\left.{\partial X \over \partial x}\right|_{t,x_2,0,0} = 
\left.{\partial X \over \partial x}\right|_{t,x_1,0,0}$$
but $x_1$ and $x_2$ are arbitrary, so $\partial X\over\partial x$ is constant
so $X(t,x,y,z,v)$ is linear with respect to $x$. Similar procedure shows, that
$X(t,x,y,z,v)$ is linear with respect to $y$, $z$ and $t$, and the same for $Y$,
$Z$ and $T$, which means, that
$$
\left( \begin{array}{c}
t' \\
x' \\
y' \\
z' \\
\end{array} \right)
=
A(v)
\left( \begin{array}{c}
t \\
x \\
y \\
z \\
\end{array} \right)
$$
where
$$
A(v)=
\left( \begin{array}{cccc}
a_{00} & a_{01} & a_{02} & a_{03} \\
a_{10} & a_{11} & a_{12} & a_{13} \\
a_{20} & a_{21} & a_{22} & a_{23} \\
a_{30} & a_{31} & a_{32} & a_{33} \\
\end{array} \right)
$$
and the coefficients $a_{\mu\nu}$ only depend on $v$. This is
the assumption \eref{ass-1}.

\subsection{Principle of relativity\label{relativity}}

The relativity principle means, that the functional form of the transformation
$A(v)$ is the same when transforming from $S'$ to $S$. The $S'$
has the speed $v$ as seen from $S$, however, the reciprocal speed of $S$ as
seen from $S'$ can be generally anything, so we denote it by $\varphi(v)$:
$$
\left( \begin{array}{c}
t \\
x \\
y \\
z \\
\end{array} \right)
=
A(\varphi(v))
\left( \begin{array}{c}
t' \\
x' \\
y' \\
z' \\
\end{array} \right)
$$
from which we get:
$$
\left( \begin{array}{c}
t \\
x \\
y \\
z \\
\end{array} \right)
=
A(\varphi(v))
A(v)
\left( \begin{array}{c}
t \\
x \\
y \\
z \\
\end{array} \right)
$$
or
$$A(\varphi(v))A(v)=\one$$
In our derivation, we assume $\varphi(v)=-v$ (and we get the
assumption \eref{ass-2}), because it is natural. However, as is shown in \cite{gorini2}, it
is not necessary, but it adds a complexity to the derivation and our motive is
not to find the weakest assumptions possible, but a reasonable set of natural
assumptions, such that the Lorentz transformation inevitably follows from them.

Now let $S''$ be moving with a speed $u$ with respect to $S'$. Then the
relativity principle requires, that transforming from $S$ to $S'$ and then to
$S''$ is the same as transforming from $S$ to $S''$ directly (with some other
speed $w$):
$$A(u)A(v) = A(w)$$
This is the assumption \eref{ass-3}.

\subsection{Isotropy\label{isotropy}}

Isotropy of space implies (among other things), that the transformation doesn't
change when we reverse the $x$-axis, i.e. that reversing the $x$-axis, applying
the transformation for the speed $v$ and reversing the $x'$-axis again is the
same as applying the transformation directly (but for some other speed $\bar
v$).  The matrix that reverses the $x$ axis is:
$$T=
\left( \begin{array}{cccc}
1 & 0 & 0 & 0 \\
0 & -1 & 0 & 0 \\
0 & 0 & 1 & 0 \\
0 & 0 & 0 & 1 \\
\end{array} \right)
$$
So the above statement means:
$$TA(v)T  = A(\bar v)$$
This is the assumption \eref{ass-4}.

The isotropy also implies, that since the only significant spacial direction is
that of the $(x,x')$-axis -- the direction of motion -- the transformation
$A(v)$ must be the same as if we first rotate about the $(x,x')$-axis,
transform and then rotate back:
$$
R(-\alpha)A(v)R(\alpha) = A(v)\label{rot}
$$
where the $R(\alpha)$ is a matrix, that rotates the system around the $x$ axis:
$$
R(\alpha) =
\left( \begin{array}{cccc}
1 & 0 & 0 & 0 \\
0 & 1 & 0 & 0 \\
0 & 0 & \cos\alpha & \sin\alpha \\
0 & 0 & -\sin\alpha & \cos\alpha \\
\end{array} \right)
$$
And this is the assumption \eref{ass-5}.

\section{Discussion}

In \ref{deriv} it is shown, that the above equations imply
$$
A(v)=
\left( \begin{array}{cccc}
{1\over\sqrt{1-Kv^2}} & -{Kv\over\sqrt{1-Kv^2}}& 0 & 0 \\
-{v\over\sqrt{1-Kv^2}}& {1\over\sqrt{1-Kv^2}} & 0 & 0 \\
0 & 0 & 1 & 0 \\
0 & 0 & 0 & 1 \\
\end{array} \right)
$$
where $K$ is a constant independent on $v$.

It can be shown
\cite{Pal} that $K<0$ is inconsistent, so we set $K={1\over c^2}$, where
$c$ is a constant, independent of the frame of reference (because $K$ is), with
a dimension of speed (possibly $c=\infty$) and we get our final formula:
$$
\left( \begin{array}{c}
t' \\
x' \\
y' \\
z' \\
\end{array} \right)
=
\left( \begin{array}{cccc}
{1\over\sqrt{1-{v^2\over c^2}}} & -{{v\over c^2}\over\sqrt{1-{v^2\over c^2}}}& 0 & 0 \\
-{v\over\sqrt{1-{v^2\over c^2}}}& {1\over\sqrt{1-{v^2\over c^2}}} & 0 & 0 \\
0 & 0 & 1 & 0 \\
0 & 0 & 0 & 1 \\
\end{array} \right)
\left( \begin{array}{c}
t \\
x \\
y \\
z \\
\end{array} \right)
$$
For $c=\infty$ we get the Galilean transformation:
$$
\left( \begin{array}{c}
t' \\
x' \\
y' \\
z' \\
\end{array} \right)
=
\left( \begin{array}{cccc}
1 & 0 & 0 & 0 \\
-v & 1 & 0 & 0 \\
0 & 0 & 1 & 0 \\
0 & 0 & 0 & 1 \\
\end{array} \right)
\left( \begin{array}{c}
t \\
x \\
y \\
z \\
\end{array} \right)
$$
For $c$ finite we get the Lorentz transformation, but
the value of $c$ is not determined by the theory and must be measured in
experiment. 

For many centuries up to around 1905, it was known from an experiment, that the
$c$ is very high or possibly infinite and it couldn't be determined at that
time, so setting $c=\infty$ was the correct answer (they didn't think
this way, but they could if they wanted and even Galileo could have derived the
special theory of relativity \cite{sen}). However now it's clear, that
the theory gives the correct results, when we set $c$ to be the speed of light
(notice however, that in general, the $c$ doesn't have to be the speed of
light).  So the speed of light axiom can actually be rephrased as: "Don't use
the Galilean transformation, because it doesn't work, and if you get some
maximum allowed speed in the theory, it is the speed of light".

\section{Invariance of the spacetime interval}

It is easy to show, that the Lorentz transformation above ($K>0$) obeys the
orthogonality relation:
$$\eta = \Lambda^T\eta\Lambda$$
where $\Lambda$ is the Lorentz transformation matrix and 
$\eta={\rm diag}(-1,1,1,1)$
is the Minkowski tensor. Written using indices:
\begin{equation}
\eta_{\alpha\beta}=\eta_{\mu\nu}\Lambda^\mu{}_\alpha\Lambda^\nu{}_\beta
\label{ort}
\end{equation}
and it can also be shown, that any transformation defined by the orthogonality
relation is either a boost (the transformation derived above), or spatial
rotations, reflections of axes or translations (see any book on the quantum
field theory, for example \cite{maggiore}). All of them are valid
transformations between $S$ and $S'$. So the
orthogonality relation can be taken as the definition of all possible
transformations between frames.

Now we define the space time interval $ds^2$ by
$$ds^2 = \eta_{\mu\nu}dx^\mu dx^\nu$$
This is invariant for all transformations defined by the orthogonality relation
\eref{ort}:
$$ds'^2 = \eta_{\mu\nu}dx'^\mu dx'^\nu=
\eta_{\mu\nu}\Lambda^\mu{}_\alpha dx^\alpha\Lambda^\nu{}_\beta dx^\beta=
\eta_{\alpha\beta}dx^\alpha dx^\beta=ds^2$$
On the other hand, all the transformations that leave the interval invariant
must be of the form \eref{ort}, because
$$ds'^2 = \eta_{\mu\nu}dx'^\mu dx'^\nu=
\eta_{\mu\nu}{dx'^\mu\over dx^\alpha} dx^\alpha{dx'^\nu\over dx^\beta}dx^\beta=
ds^2=\eta_{\alpha\beta}dx^\alpha dx^\beta$$
This is true for all $dx^\alpha$ and $dx^\beta$, so we get:
\begin{equation}
\eta_{\mu\nu}{dx'^\mu\over dx^\alpha}{dx'^\nu\over dx^\beta}=
\eta_{\alpha\beta}\label{ort2}
\end{equation}
It can also be shown \cite{weinberg} that this equation implies:
$${d^2x'^\mu\over dx^\rho dx^\alpha}=0$$
But then 
$${dx'^\mu\over dx^\alpha} = \Lambda^\mu{}_\alpha$$
are constants (depending only on $v$) and \eref{ort2} are the orthogonality
relations \eref{ort}. In other words, the orthogonality relations are
equivalent to the invariance of the interval. 

So the starting point to the special theory of relativity can be any of these
(all of them are equivalent, as shown in this paragraph):
\begin{itemize}
\item homogeneity, isotropy, the principle of relativity and the requirement,
that we don't want the Galileo transformation
\item the orthogonality relation
\item invariance of the spacetime interval
\end{itemize}

\section{Conclusion}

We showed from the homogeneity, isotropy and the principle of
relativity that the only possible transformations between $S$ and $S'$
are either the Galileo or Lorentz transformation, but nothing else. Contrary to
other texts, we first wrote explicit equations and then only used a pure
algebra to derive our result.

\section{Acknowledgements}

I thank Oldrich Semerak for listening to my arguments and for showing me some
references and books. I also thank to A. Fejfar, K. Vyborny and O.
Semerak for reading the manuscript. This research was partly supported by the
LC06040 research center project.

\appendix

\section{Derivation of the Lorentz transformation}\label{deriv}

From \eref{ass-5} we get by multiplying by $R(\alpha)$ from left:
$$R(\alpha)A(v) = A(v)R(\alpha)$$
This must hold for any $\alpha$ and in \ref{rotation} it is shown, that
$$
A(v)=
\left( \begin{array}{cc}
A_1 & 0 \\
0 & kP(\theta) \\
\end{array} \right)
$$
where
$$
kP(\theta)=
\left( \begin{array}{cc}
k\cos\theta & k\sin\theta \\
-k\sin\theta & k\cos\theta \\
\end{array} \right)
$$
for some values of the parameters $k(v)$ and $\theta(v)$, that are functions of
$v$. However, from \eref{ass-10} we get (for all $v$ and $z$):
$$
\left( \begin{array}{c}
0 \\
z' 
\end{array} \right)
=
\left( \begin{array}{cc}
k\cos\theta & k\sin\theta \\
-k\sin\theta & k\cos\theta 
\end{array} \right)
\left( \begin{array}{c}
0 \\
z
\end{array} \right)
$$
From which $zk\sin\theta=0$ for all $z$, so $k\sin\theta=0$ and that implies
$$
kP(\theta)=
\left( \begin{array}{cc}
k(v) & 0 \\
0 & k(v) \\
\end{array} \right)
$$
This $k(v)$ can be positive, negative or zero.
So now $A(v)$ has this form:
$$
\left( \begin{array}{c}
t' \\
x' \\
y' \\
z' \\
\end{array} \right)
=
\left( \begin{array}{cccc}
D(v) & C(v) & 0 & 0 \\
B(v) & A(v) & 0 & 0 \\
0 & 0 & E(v) & 0 \\
0 & 0 & 0 & E(v) \\
\end{array} \right)
\left( \begin{array}{c}
t \\
x \\
y \\
z \\
\end{array} \right)
$$
where the constants $A$, $B$, $C$, $D$, $E$ only depend on $v$, the direct
velocity, and from \eref{ass-9} we get 
$0=A(v) vt+B(v) t$ so
\begin{equation}
v=-{B(v)\over A(v)}\label{vBA}
\end{equation}
In other words, we can always determine the direct speed $v$ 
from the matrix elements. From \eref{ass-4} we get
$$
TA(v)T = 
\left( \begin{array}{cccc}
D(v) & -C(v) & 0 & 0 \\
-B(v) & A(v) & 0 & 0 \\
0 & 0 & E(v) & 0 \\
0 & 0 & 0 & E(v) \\
\end{array} \right)
$$
$$
A(\bar v) = 
\left( \begin{array}{cccc}
D(\bar v) & C(\bar v) & 0 & 0 \\
B(\bar v) & A(\bar v) & 0 & 0 \\
0 & 0 & E(\bar v) & 0 \\
0 & 0 & 0 & E(\bar v) \\
\end{array} \right)
$$
Comparing the two matrices we see that $B(\bar v) = -B(v)$ and 
$A(\bar v)=A(v)$. However, from \eref{vBA} we have
$\bar v = - {B(\bar v) \over A(\bar v)}$ and $v=-{B(v) \over A(v)}$, but then
$\bar v = - {B(\bar v) \over A(\bar v)} = {B(v) \over A(v)} = -v$
and we get these relations by comparing the matrix elements of the two matrices:
\begin{eqnarray}
A(-v) &= A(v)\label{sym1}\\
B(-v) &= -B(v)\\
C(-v) &= -C(v)\\
D(-v) &= D(v)\\
E(-v) &= E(v)\label{sym2}
\end{eqnarray}
Using \eref{ass-2} and the symmetries \eref{sym1} -- \eref{sym2} we get:
$$
A(-v)A(v)=
\left( \begin{array}{cccc}
D(v) & -C(v) & 0 & 0 \\
-B(v) & A(v) & 0 & 0 \\
0 & 0 & E(v) & 0 \\
0 & 0 & 0 & E(v) \\
\end{array} \right)
\left( \begin{array}{cccc}
D(v) & C(v) & 0 & 0 \\
B(v) & A(v) & 0 & 0 \\
0 & 0 & E(v) & 0 \\
0 & 0 & 0 & E(v) \\
\end{array} \right)
=
\one
$$
multiplying:
$$
\left( \begin{array}{cccc}
D^2-BC & C(D-A) & 0 & 0 \\
B(A-D) & A^2-BC & 0 & 0 \\
0 & 0 & E^2 & 0 \\
0 & 0 & 0 & E^2 \\
\end{array} \right)
=
\one
$$
or
\begin{eqnarray}
A^2 - BC &= 1\label{cA}\\
B(A-D)   &= 0\label{cB}\\
D^2 - BC &= 1\\
C(A-D)   &= 0\\
E^2      &= 1 \label{cE}
\end{eqnarray}
From \eref{cE} we get $E(v)=\pm1$, but from \eref{ass-7} we have $E(0)=1$ so
$E(v) = 1$ (of course we require that matrix elements are continuous).  

If for some $v$ the $A(v)\neq D(v)$, then $B(v)=0$ from \eref{cB}, thus
$A(v)=\pm1$ from \eref{cA} and from \eref{vBA} we get $v=-{0\over\pm1}=0$,
which means that
$A(0)\neq D(0)$, but that is a contradiction with \eref{ass-7}, that asserts
$A(0)=D(0)=1$.

So we must have $A(v)=D(v)$ for all $v$, then from \eref{cA} we get
$C(v)={A^2(v)-1\over B(v)}$ and from \eref{vBA} follows $B(v)=-vA(v)$:
$$
\left( \begin{array}{c}
t' \\
x' \\
y' \\
z' \\
\end{array} \right)
=
\left( \begin{array}{cccc}
A & -{A^2-1\over vA} & 0 & 0 \\
-vA & A & 0 & 0 \\
0 & 0 & 1 & 0 \\
0 & 0 & 0 & 1 \\
\end{array} \right)
\left( \begin{array}{c}
t \\
x \\
y \\
z \\
\end{array} \right)
$$
where $A(v)$ is an unknown function of $v$, except that $A(0)=1$ (follows from
\eref{ass-7}).
Now we use \eref{ass-3}:
$$
A(u)A(v)
=
\left( \begin{array}{cccc}
A_u & -{A_u^2-1\over uA_u} & 0 & 0 \\
-uA_u & A_u & 0 & 0 \\
0 & 0 & 1 & 0 \\
0 & 0 & 0 & 1 \\
\end{array} \right)
\left( \begin{array}{cccc}
A_v & -{A_v^2-1\over vA_v} & 0 & 0 \\
-vA_v & A_v & 0 & 0 \\
0 & 0 & 1 & 0 \\
0 & 0 & 0 & 1 \\
\end{array} \right)
=A(w)
$$
Multiplying the matrices:
$$
A(u)A(v)
=
\left( \begin{array}{cccc}
A_uA_v + (A_u^2-1){vA_v\over uA_u} & \dots & 0 & 0 \\
\dots & A_uA_v + (A_u^2-1){uA_u\over vA_v} & 0 & 0 \\
0 & 0 & 1 & 0 \\
0 & 0 & 0 & 1 \\
\end{array} \right)
$$
and
$$
A(w)
=
\left( \begin{array}{cccc}
A_w & -{A_w^2-1\over wA_w} & 0 & 0 \\
-wA_w & A_w & 0 & 0 \\
0 & 0 & 1 & 0 \\
0 & 0 & 0 & 1 \\
\end{array} \right)
$$
so comparing the two expressions for $A_w$ (the first and the second diagonal
element) we get:
$${A_v^2-1\over v^2A_v^2} = {A_u^2-1\over u^2A_u^2}$$
where the left hand side only depends on $v$, the right hand side only on $u$,
thus both sides are equal to a constant $K$, that is independent of the frame
of reference, because it doesn't depend on the coordinates or $v$, so
we get (remember A(0)=1, so we take the positive square root)
$$A_v = {1\over\sqrt{1-Kv^2}}$$ 
and we arrive at the expression for the
transformation between $S$ and $S'$:
$$
\left( \begin{array}{c}
t' \\
x' \\
y' \\
z' \\
\end{array} \right)
=
\left( \begin{array}{cccc}
{1\over\sqrt{1-Kv^2}} & -{Kv\over\sqrt{1-Kv^2}}& 0 & 0 \\
-{v\over\sqrt{1-Kv^2}}& {1\over\sqrt{1-Kv^2}} & 0 & 0 \\
0 & 0 & 1 & 0 \\
0 & 0 & 0 & 1 \\
\end{array} \right)
\left( \begin{array}{c}
t \\
x \\
y \\
z \\
\end{array} \right)
$$

\section{Rotations}\label{rotation}

For each $\alpha$, we have:
$$R(\alpha)A(v) = A(v)R(\alpha)$$
where
$$
R(\alpha) =
\left( \begin{array}{cccc}
1 & 0 & 0 & 0 \\
0 & 1 & 0 & 0 \\
0 & 0 & \cos\alpha & \sin\alpha \\
0 & 0 & -\sin\alpha & \cos\alpha \\
\end{array} \right)
=
\left( \begin{array}{cc}
\one & 0 \\
0 & P(\alpha) \\
\end{array} \right)
$$
$$
P(\alpha) =
\left( \begin{array}{cc}
\cos\alpha & \sin\alpha \\
-\sin\alpha & \cos\alpha \\
\end{array} \right)
=\one\cos\alpha+i\sigma_2\sin\alpha=e^{i\alpha\sigma_2}
$$
$$
A(v)=
\left( \begin{array}{cc}
A_1 & A_2 \\
A_3 & A_4 \\
\end{array} \right)
$$
and the $\sigma_1$, $\sigma_2$ and $\sigma_3$ are the Pauli matrices.
Then
$$R(\alpha)A(v) - A(v)R(\alpha)=
\left( \begin{array}{cc}
\one & 0 \\
0 & P(\alpha) \\
\end{array} \right)
\left( \begin{array}{cc}
A_1 & A_2 \\
A_3 & A_4 \\
\end{array} \right)
-
\left( \begin{array}{cc}
A_1 & A_2 \\
A_3 & A_4 \\
\end{array} \right)
\left( \begin{array}{cc}
\one & 0 \\
0 & P(\alpha) \\
\end{array} \right)
=
$$
$$=
\left( \begin{array}{cc}
0 & A_2(\one-P(\alpha)) \\
(P(\alpha)-\one)A_3 & P(\alpha)A_4-A_4P(\alpha) \\
\end{array} \right)
=0
$$
The parameter $\alpha$ is arbitrary, so $A_2=A_3=0$ and (we set 
$A_4=a_0\one+a_1\sigma_1+a_2\sigma_2+a_3\sigma_3$)
$$P(\alpha)A_4-A_4P(\alpha)=e^{i\alpha\sigma_2}
(a_0+a_1\sigma_1+a_2\sigma_2+a_3\sigma_3)-(a_0+a_1\sigma_1+a_2\sigma_2+a_3\sigma_3)
e^{i\alpha\sigma_2}=$$
$$=e^{i\alpha\sigma_2}(a_1\sigma_1+a_3\sigma_3)-(a_1\sigma_1+a_3\sigma_3)
e^{i\alpha\sigma_2}
=i\sin\alpha\left(\sigma_2(a_1\sigma_1+a_3\sigma_3)-(a_1\sigma_1+a_3\sigma_3)
\sigma_2\right)=$$
$$=2\sin\alpha(a_1\sigma_3-a_3\sigma_1)=0$$
Multiplying by $\sigma_3$ from the left and taking a trace we get
$$\Tr2\sigma_3\sin\alpha(a_1\sigma_3-a_3\sigma_1)=
2\sin\alpha(a_1\Tr\one-ia_3\Tr\sigma_2)=0$$
but $\Tr\sigma_2=0$ and $\Tr\one=2$ so $a_1=0$. Similarly $a_3=0$. So
$$A_4=a_0+a_2\sigma_2=ke^{i\theta\sigma_2}=kP(\theta)$$
where $k=\sqrt{a_0^2+a_2^2}$, $\cos\theta={a_0\over k}$ and 
$\sin\theta={a_2\over k}$. So the matrix $A(v)$ can always be written as:
$$
A(v)=
\left( \begin{array}{cc}
A_1 & 0 \\
0 & kP(\theta) \\
\end{array} \right)
$$
for some values of the parameters $k(v)$ and $\theta(v)$, that are functions of
$v$. Note, that if we rotate the axes before doing the transformation:
$$
A(v)R(\alpha)=
\left( \begin{array}{cc}
A_1 & 0 \\
0 & kP(\theta) \\
\end{array} \right)
\left( \begin{array}{cc}
\one & 0 \\
0 & P(\alpha) \\
\end{array} \right)
=
\left( \begin{array}{cc}
A_1 & 0 \\
0 & kP(\theta)P(\alpha) \\
\end{array} \right)
$$
We see that by rotating around the $x$-axis by the angle $\alpha=-\theta$,
we get
$$
A(v)R(-\theta)=
\left( \begin{array}{cc}
A_1 & 0 \\
0 & k\one \\
\end{array} \right)
$$
Geometrically this means, that the $A(v)R(-\theta)$ doesn't rotate the
$y$ and $z$ axes (only scales them by a factor of $k$).

\section*{References}
\bibliography{citations}

\end{document}